\documentstyle[12pt]{article}
\begin{document}
\title {Ultra-Violet Behavior of Bosonic Quantum Membranes}
\author {Michio Kaku}
\date {Physics Dept., City College of New York\\
New York, N.Y. 10031, USA}
\maketitle
\centerline {ABSTRACT}
We treat the action for a bosonic membrane as a sigma model,
and then compute quantum corrections by integrating out higher
membrane modes. As in string theory, where the equations of motion
of Einstein's theory emerges by setting $\beta = 0$,
we find that, with certain assumptions,
we can recover the equations of motion for the background fields,
i.e. $R _ { \mu \nu } + ( 1/4) F _ {\mu  \alpha \beta \gamma \mu} 
F _ \nu ^ { \alpha\beta \gamma} = 0$ 
and $D ^ \alpha F _ { \alpha \beta \gamma} = 0$.
for the membrane case. Although the
membrane theory is non-renormalizable on the world volume
by power counting, the investigation of the
ultra-violet behavior of
membranes may give us insight into the supersymmetric
case, where
we hope to obtain
higher order M-theory corrections to 11 dimensional supergravity.

\section{Introduction}
At present, little  is known about the action for M-theory [1,2], other
than that it contains 11 dimensional supergravity in the 
low energy limit. Higher order corrections to 11D supergravity [3] are
unknown.
In this series of papers, we hope to compute these higher order
corrections.

In string theory, the usual 10D supergravity action is derived
by treating the original string action as a sigma model
and them integrating the higher modes. By setting
$\beta = 0$ via conformal invariance, we then
obtain the equations of motion of 10D supergravity, plus
higher order corrections to any order [4].

We would like to apply this same general technique to
M-theory, treating the 11D supermembrane action [5]
as a sigma model in order to compute
the higher order corrections to M-theory.
There are, of course, several obstacles to performing such
a procedure. 

First, by power counting, the membrane theory in higher dimensions is 
non-renormalizable on the world volume. We find that the degree of
divergence for any
N-point function
can be made arbitrarily high by adding 
higher vertex corrections, thereby rending invalid the standard 
renormalization technique.

Second, there are problems with the quantization of
supermembranes, i.e. they are quantum mechanically unstable [6].
(This instability was the original reason why many
abandoned supermembrane theory soon after it was
formulated. However, it may be possible to re-interpret
this instability in terms of 0-branes in matrix models [7].)

Third, the precise relationship between membranes and M-theory
is also not known. In particular, M-theory also contains five-branes,
and perhaps higher order corrections to membranes as well.

Our philosophy, however, will be to investigate 
the first problem.
Although the model is superficially non-renormalizable by
power counting methods, it may possess enough
symmetry to eliminate large classes of diverges.
For example, there is no counterpart to the
$\beta = 0$ equation for membranes, because there is
no conformal symmetry on the world volume. However,
in a later paper we will show that supersymmetry
will in fact set the  analogous supermembrane term
to zero because of the super Bianchi identities.
Thus, supersymmetry (which demands compatibility with
the 11D supergravity background equations) is sufficient
to render the theory one-loop renormalizable.

Our ultimate goal is to show whether or not
supersymmetry is sufficient to
kill the divergences of the supermembrane
theory to all orders.
This, in turn, would allow us to compute the higher order
corrections to 11D supergravity in the M-theory action.
If a recursion relation can be written for the
higher order corrections, then we may be able
to make statements concerning the entire theory,
to all orders.

However, even if this ultimate goal is not realized,
we expect to find interesting surprises. 
For example, we will show that, unlike the string case,
one needs both one-loop and two-loop graphs in order to
derive the standard equations of motion for the
graviton and anti-symmetric tensor field.
In the same way that the non-renormalizable
four-fermion theory or the massive vector meson
theory proved to be interesting
laboratories for particle physics, it may turn out
that supermembrane actions, even if they are
inherently non-renormalizable, may be an interesting
laboratory for M-theory.

\section {Riemann Normal Co-ordinates}

Our starting point is the bosonic membrane action:

\begin {equation}
L _ 1 = 
{ 1 \over 2 \alpha } 
{\sqrt \gamma } \gamma ^ { ij}
g _ { \mu \nu } \partial _ i \phi ^ \mu
\partial _ j  \phi ^ \nu 
\end {equation}
where $g _ { \mu \nu}$ is the space-time metric,
where Greek letters $\mu, \nu , \alpha = 0, 1 , 2 ... 10$,
where $\gamma ^ { ij}$ is the
metric on the three-dimensional world volume,
where Roman letters $i,j , k = 1,2,3$, and where $\phi ^ \mu$ is the
membrane co-ordinate. 

To this action, we add a contribution from the
anti-symmetric field:

\begin {equation}
L _ 2 = 
\beta \epsilon ^ {ijk } A _ { \mu \nu \lambda }
\partial _ i \phi ^ \mu
\partial _ j \phi ^ \nu
\partial _ k \phi ^ \lambda
\end {equation}
which is found in the bosonic part of the
supermembrane action.
The total action is then $L _ T = L _ 1+ L _ 2$,
with $\alpha$ and $\beta$ being
two coupling constants.

Notice that the action of this theory is gauge invariant
under the transformation:

\begin {equation}
\delta A _ { \mu \nu \lambda }
= \partial _ \mu \Lambda _ { \nu \lambda } + ...
\end {equation}

Notice that the action also contains a 
world volume metric $\gamma ^ { ij }$.
In the usual string action, this metric can be
eliminated entirely via a gauge choice and
a conformal transformation. However, in the
covariant membrane case, we cannot eliminate
all the degrees of freedom of the
non-propagating world volume metric. 
Instead, we will simply treat the
metric $\gamma ^ { ij }$
as a classical background field.
This means that we will have to keep
$\gamma ^ {i j} $ arbitrary and
quantize the theory
on a classical curved world volume.

Next, we wish to power expand this action using the
background field method applied to sigma models, using
Riemann normal co-ordinates [6].
Let the space-time variable $\phi ^ \mu (\tau)$ 
obey a standard geodesic equation:

\begin{equation}
{ d ^ 2 \phi ^ \mu \over d \tau ^ 2 } 
+ 
\Gamma _ { \rho \sigma } ^ \mu
{ d \phi ^ \rho  \over d \tau}
{ d \phi ^ \sigma \over d \tau}
= 0
\end {equation}

Now expand the membrane co-ordinate $\phi ^\mu$
around a classical configuration $\phi _ { \rm cl }^ \mu$:

\begin {equation}
\phi ^ \mu = \phi _ {\rm cl } ^ \mu +
\pi ^ \mu
\end {equation}
where $\pi ^\mu$ is the quantum correction to the classical
configuration.
Now power expand $\pi ^ \mu$ in terms of 
$\xi^ \mu$:

\begin {equation}
\pi ^ \mu = 
\xi ^ \mu 
- { 1 \over 2 } 
\Gamma _ { \rho \sigma } ^ \mu
\xi ^ \rho \xi ^ \sigma 
- { 1 \over 3!}
\Gamma _ { \rho \sigma \lambda } ^ \mu
\xi ^ \rho
\xi ^ \sigma 
\xi ^ \lambda
...
\end {equation}
The various co-efficients in this power expansion can be
laboriously computed by inserting the expression back
into the geodesic equation.
For example, we find that $\Gamma _ { \mu \nu } ^ \lambda$
is the usual Christoffel symbol, and:

\begin {equation}
\Gamma _ { \rho \sigma \lambda } ^ \mu
=
\partial _ \rho
\Gamma _ { \sigma \lambda } ^ \mu
- 
\Gamma _ {\rho \sigma} ^ \alpha
\Gamma _ { \alpha \lambda } ^ \mu
-
\Gamma _ {\rho \lambda} ^ \alpha
\Gamma _ { \alpha \sigma} ^ \mu
\end {equation}
In general, the higher coefficients are 
equal to:

\begin {equation}
\Gamma _ { \mu _ 1 \mu _ 2 ...  \mu _ n \alpha \beta } ^ \nu
= D _ { \mu _ n } \cdots D _ { \mu _ 1 } 
\Gamma _ { \alpha \beta } ^ \nu
\end {equation}
where we take the covariant derivatives only with respect to the
lower indices.

Our goal is now to power expand the Lagrangian $L _ 1 + L _2$
in terms of $\xi ^ \mu$, and then integrate out
$\xi ^\mu$ from the action.
This will give us a series of potentially divergent
graphs, whose structure we wish to examine.

In general, this power expansion becomes prohibitively
difficult as we progress to higher and higher orders,
so we will instead use the formalism introduced by
Mukhi [9].

One reason why this expansion is unwieldy is because
the standard Taylor expansion is non-covariant.
If we have a function $I$ and power expand it, we find:

\begin {equation}
I =
\sum _ { n=0 } ^ \infty
I ^ { (n)}
\end {equation}
where:

\begin {equation}
I ^ { (n)} =
{ 1 \over n!}
\int d x _ 1 \xi ^ { \mu _ 1} \partial _ { \mu _ 1 } 
^ { x _ 1 }
\int d x _ 2 \xi ^ { \mu _ 2} \partial _ { \mu _ 2} 
^ { x _ 2 }
\cdots
\int d x _ n \xi ^ { \mu _ n} \partial _ { \mu _ n } 
^ { x _ n }
I
\end {equation}
where the co-ordinates on the three dimensional
world volume are given by $x _ i$,
and where $\partial _ i ^ x$ is a functional derivative:
\begin {equation}
\partial _ \mu ^ x = 
{\delta 
\over
\delta \phi ^ \mu ( x ) }
\end {equation}

Clearly, iterating the operator:
\begin {equation}
\int d x \, \xi ^ \mu ( x ) \partial _ \mu ^ x
\end {equation}
yields non-covariant results.

Let us define instead the operator $\Delta$:

\begin {equation}
\Delta = \int dx \, \xi ^ \mu ( x ) D _ \mu
\end {equation}
where $D _ \mu$ is a functional covariant derivative.
For example:

\begin {equation}
D _ \mu A ^ \nu [ \phi (y )]
=
\left [
\partial _ \mu A ^ \nu ( \phi ( x ) )
+ \Gamma _ { \mu \lambda} ^ \nu 
( \phi ( x ) ) A ^ \lambda ( \phi ( x ) ) \right ]
\delta ^ 3 ( x - y )
\end {equation}

Then we can power expand the Lagrangian as follows:

\begin {equation}
L = \sum _ { n=0 } ^ \infty L ^ { (n)}
\end {equation}
where:
\begin {equation}
L ^ { (n)} =
{ 1 \over n !}
\Delta ^ n L
\end {equation}

To perform the power expansion, we derive the following
identities:

\begin {eqnarray}
\Delta \xi ^ \mu &=& 0
\cr
\Delta ( \partial _ i \phi ^ \mu ) & = & 
D _ i \xi ^ \mu
\cr
\Delta ( D _ i \xi ^ \mu ) & = & 
R _ { \nu \rho \sigma } ^ \mu
\xi ^ \nu
\xi ^ \rho
\partial _ i \phi ^ \sigma
\cr
\Delta T _ { \mu _ 1 \mu _ 2 ... } & = & 
\xi ^ \rho D _ { \rho } T _ { \mu _ 1 \mu _ 2 ... }
\end {eqnarray}
where $ T _ { \mu _ 1 \mu _ 2 ... }$ is an arbitrary
tensor, and:

\begin {equation}
D _ i \xi ^ \mu = 
\partial _ i \xi ^ \mu +
\Gamma _ { \rho \sigma } ^ \mu 
\xi ^ \rho \partial _ i \phi ^ \sigma
\end {equation}
and:
\begin {equation}
R _ { \nu \rho \sigma } ^ \mu 
=
\partial _ \rho \Gamma _ { \nu \sigma } ^ \mu
- \partial _ \sigma \Gamma _ { \nu \rho } ^ \mu
+
\Gamma _ { \nu \sigma } ^ \lambda
\Gamma _ { \lambda \rho } ^ \mu
-
\Gamma _ { \nu \rho } ^ \lambda
\Gamma _ { \lambda \sigma } ^ \mu
\end {equation}

Now let us power expand the original action in terms of
$\xi$. Let us replace $\phi _ { \rm cl} ^ \mu$ with the
symbol $\phi ^ \mu$.

We find:
\begin {eqnarray}
L _ 1 ^ { (0)} & = &
{ 1\over 2 \alpha }
\sqrt \gamma \gamma ^ { ij } 
g _ { \mu\nu}
\partial _ i \phi ^ \mu
\partial _ j \phi ^ \nu 
\cr
L _ 1 ^ { (1)} 
&= &
{ 1 \over \alpha } {\sqrt \gamma }
\gamma ^ { ij} 
g _ { \mu \nu }
\partial _ i \phi ^ \mu
D _ j \xi ^ { \nu } 
\cr
L _ 1 ^ { (2)} 
&=&
{ 1 \over 2 \alpha } 
{\sqrt \gamma }
\gamma ^ { ij} 
R _ { \mu \nu \sigma \rho }
\partial _ i \phi ^ \mu
\partial _ j \phi ^ \sigma 
\xi ^ \nu \xi ^ \rho
+ 
{ 1 \over 2 \alpha } 
{\sqrt \gamma }
\gamma ^ { ij} 
g _ { \mu \nu } 
D _ i \xi ^ \mu
D _ j \xi ^ \nu
\cr
L _ 1 ^ { (3)} 
&=&
 { 1 \over 6 \alpha } 
{\sqrt \gamma }
\gamma ^ { ij} 
R _ { \mu \nu \rho \sigma ; \lambda } 
\partial _ i \phi ^ \mu
\partial _ j \phi ^ \sigma
\xi ^ \nu 
\xi ^ \rho
\xi ^ \lambda
+ 
{ 2 \over 3 \alpha }
{\sqrt \gamma}
\gamma ^ { ij }
R _ { \mu \nu \rho \sigma }
\partial _ i \phi ^ \mu 
D _ j \xi ^ \sigma
\xi ^ \nu
\xi ^ \rho
\cr
L _ 1 ^ { (4)}
&=&
{ 1 \over 24\alpha } 
{\sqrt \gamma }
\gamma ^ { ij} 
\left [
R _ { \mu \nu \rho \sigma ; 
\pi \kappa } 
+ 4 R _ { \nu \rho \mu } ^ \lambda
R _ { \lambda \pi \kappa \sigma } \right]
\partial _ i \phi ^ \mu
\partial _ j \phi ^ \sigma
\xi ^ \nu
\xi ^ \sigma
\xi ^ \pi
\xi ^ \kappa
\cr
&+ &
{ 1 \over 4 \alpha }
{\sqrt \gamma }
\gamma ^ { ij} 
R _ { \mu \rho \sigma \nu ; \pi } 
\partial _ i \phi ^ \mu D _ j \xi ^ \nu
\xi ^ \rho \xi ^ \sigma \xi ^ \pi 
+
{ 1 \over 6 \alpha } 
{\sqrt \gamma }
\gamma ^ { ij} 
R _ { \mu \sigma \rho \nu }
D _ i \xi ^ \mu D _ i \xi ^ \nu
\xi ^ \rho
\xi ^ \sigma 
\cr
L _ 1 ^ { (5)} & = &
{ 1 \over 120 \alpha }
{\sqrt \gamma } \gamma ^ { ij }
\left( R  _ { \mu \nu \alpha \beta ;
\gamma \delta \lambda}
+ 14 R _ { \alpha \beta \mu ; \gamma } ^ \epsilon
R _ { \epsilon \delta \lambda \nu} \right)
\partial _ i \phi ^ \mu 
\partial _ j \phi ^ \nu
\xi ^ \alpha
\xi ^ \beta
\xi ^ \gamma
\xi ^ \delta
\xi ^ \lambda
\cr
& + &
{ 1 \over 15 \alpha}
{\sqrt \gamma } \gamma ^ { ij }
\left( R _ {\mu \alpha \beta \nu ; \gamma \delta }
+
2 R _ { \alpha \beta \mu } ^ \pi
R _ { \pi \gamma \delta \nu} \right) 
\partial _ i \phi ^ \mu
D _ j \xi ^ \nu
\xi ^ \alpha
\xi ^ \beta
\xi ^ \gamma
\xi ^ \delta
\cr
&+& 
{ 1 \over 12 \alpha}
{\sqrt \gamma } \gamma ^ { ij }
R _ { \mu \alpha \beta \nu ; \gamma}
D _ i  \xi ^ \mu
D _ j \xi ^ \nu
\xi ^ \alpha
\xi ^ \beta
\xi ^ \gamma
\cr
L _ 1 ^ { (6)}
& = &
{ 1 \over 720 \alpha}
{\sqrt \gamma } \gamma ^ { ij }
\left ( R _ { \mu \alpha \beta \nu ; \gamma \delta \epsilon
\lambda}
+ 22 R _ { \mu \alpha \beta \pi ; \gamma \delta } 
R _ { \epsilon \lambda \nu} ^ \pi
+ 14 R _ { \alpha \beta \mu ; \gamma } ^ \pi
R _ { \pi \delta \epsilon \nu ; \lambda} \right.
\cr
& + & 
\left.
16 R _ { \alpha \beta \mu } ^ \pi
R _ { \pi \gamma \delta \rho}
R _ { \epsilon \lambda \nu } ^ \rho 
\right)
\partial _ i \phi ^ \mu
\partial _ i \phi ^ \nu
\xi ^ \alpha
\xi ^ \beta
\xi ^ \gamma
\xi ^ \delta
\xi ^ \epsilon
\xi ^ \lambda
\cr
&+&
{ 1 \over 72 \alpha}
{\sqrt \gamma } \gamma ^ { ij }
\left(  R _ { \mu \alpha \beta \nu ; \gamma
\delta \epsilon}
+ 4 R _ { \alpha \beta \mu; \gamma}
^ \pi
R _ { \pi \delta \epsilon \nu }
+
R _ { \alpha \beta \nu ; \gamma } ^ \pi
R _ { \pi \delta \epsilon \mu } \right)
\partial _ i \phi ^ \mu
D _ i \xi ^ \nu
\xi ^ \alpha
\xi ^ \beta
\xi ^ \gamma
\xi ^ \delta
\xi ^ \epsilon
\cr
& +  &
{ 1 \over 40 \alpha}
{\sqrt \gamma } \gamma ^ { ij }
\left( R _ { \mu \alpha \beta \nu ; \gamma \delta }
+ { 8 \over 9}
R _ { \alpha \beta \mu } ^ \pi
R _ { \pi \gamma \delta \nu } \right)
D _ i \xi ^ \mu
D _ j \xi ^ \nu
\xi ^ \alpha
\xi ^ \beta
\xi ^ \gamma
\xi ^ \delta
\end {eqnarray}

The power expansion of the Lagrangian involving the 
anti-symmetric field is given by:

\begin {eqnarray}
L _ 2 ^ { (0)}
&=&
\beta {\sqrt \gamma }
\epsilon ^ { ij  k } 
A _ { \mu \nu \lambda }
\partial _ i \phi ^ \mu
\partial _ j \phi ^ \nu
\partial _ k \phi ^ \lambda
\cr
L _ 2 ^ { (1)}
& =&
3 \beta 
\epsilon ^ {ijk } 
A _ { \mu \nu \lambda }
D _ i \xi ^ \mu
\partial _ j \phi ^ \nu
\partial _ k \phi ^ \lambda
+
\beta \epsilon ^ {ijk } 
D _ \rho A _ { \mu \nu \lambda}
\xi ^ \rho \partial _ i \phi ^ \mu
\partial _ j \phi ^ \nu
\partial _ k \phi ^ \lambda
\end {eqnarray}

Since the original action was gauge invariant in the
anti-symmetric field, we wish to preserve this symmetry,
so let us re-write $L _ 2 ^ { (1)}$
by introducing the tensor:

\begin {equation}
F _ { \sigma \mu \nu \lambda } =
D _ \sigma A _ { \mu \nu \lambda } 
- D _ \mu A _ { \nu \lambda \sigma }
+ D _ \nu A _ { \lambda \sigma \mu }
- D _ \lambda A _ { \sigma \mu \nu }
\end {equation}
Because of gauge invariant, all subsequent terms will involve
this covariant tensor.

Then we can write:

\begin {equation}
L _ 2 ^ { (1)} =
\beta \epsilon ^ {ijk } 
F _ { \sigma \mu \nu \lambda }
\partial _ i \phi ^ \mu
\partial _ j \phi ^ \nu
\partial _ k \phi ^ \lambda
\xi ^ \sigma
\end {equation}

\begin {equation}
L _ 2 ^ { (2)}
=
{ \beta  \over 2 }
\epsilon ^ {ijk } 
\left (
3 F _ { \sigma \mu \nu \lambda } 
D _ i  \xi ^ \mu \partial _ j \phi ^ \nu
\partial _ k \phi ^ \lambda 
\xi ^ \sigma
+
D _ \rho F _ { \sigma \mu \nu \lambda }
\partial _ i \phi ^ \mu
\partial _ j \phi ^ \nu
\partial _ k \phi ^ \lambda
\xi ^ \rho
\xi ^ \sigma 
\right )
\end {equation}

\begin {eqnarray}
L _ 2 ^ { (3)}
&=&
{ \beta  \over 6 } 
\epsilon ^ { ijk }
\left (
6 F _ { \sigma \mu \nu \lambda}
D _ i \xi ^ \mu
D _ j \xi ^ \nu
\partial _ k \phi ^ \lambda
\xi ^ \sigma
+
6 D _ \rho F _ { \sigma \mu \nu \lambda}
D _ i \xi ^ \mu
\partial _ j \phi ^ \nu
\partial _ k \phi ^ \lambda
\xi ^ \rho \xi ^ \sigma \right.
\cr
& + &
3
R _ { \alpha \beta \gamma } ^ \mu
\partial _ i \phi ^ \gamma
F _ { \sigma \mu \nu \lambda }
\xi ^ \alpha
\xi ^ \beta 
\xi ^ \sigma
\partial _ j \phi ^ \nu
\partial _ k \phi ^ \lambda
\cr
& + &
\left. D _ \pi D _ \rho F _ { \sigma \mu \nu \lambda}
\partial _ i \phi ^ \mu
\partial _ j \phi ^ \nu
\partial _ k \phi ^ \lambda
\xi ^ \rho \xi ^ \sigma
\xi ^ \pi
\right )
\end {eqnarray}

\begin {eqnarray}
L _ 2 ^ { (4)}
&=&
{ \beta  \over 4!}
\epsilon ^ { ijk}
\left \{
6 F _ { \sigma \mu \nu \lambda}
D _ i \xi ^ \mu
D _ j \xi ^ \nu
D _ k \xi ^ \lambda
\xi ^ \sigma \right.
\cr
&+&
18 R _ { \alpha \beta \gamma } ^ \mu
F _ { \sigma \mu \nu \lambda}
\partial  _ i \phi ^ \gamma 
D _ j \xi ^ \nu
\partial _ k \phi ^ \lambda
\xi ^ \alpha
\xi ^ \beta
\xi ^ \sigma
\cr
&+&
18
D _ \rho F _ { \sigma \mu \nu \lambda}
D _ i \xi ^ \mu
D _ j \xi ^ \nu
\partial _ k \phi ^ \lambda
\xi ^ \sigma
\xi ^ \rho
\cr
& + &
9
D _ \pi D _ \rho F _ { \sigma \mu \nu \lambda}
D _ i \xi ^ \mu
\partial _ j \phi ^ \nu
\partial _ k \phi ^ \lambda
\xi ^ \sigma \xi ^ \rho \xi ^ \pi
\cr
& + & 
9
R _ { \alpha \beta \gamma } ^ \mu
D _ \rho
F _ { \sigma \mu \nu \lambda}
\partial _ i \phi ^ \gamma 
\partial _ j \phi ^ \nu
\partial _ k \phi ^ \lambda
\xi ^ \alpha \xi ^ \beta
\xi  ^ \rho 
\xi ^ \sigma
\cr
& + &
D _ \kappa D _ \pi D _ \rho
F  _ { \sigma \mu \nu \lambda}
\partial _ i \phi ^ \mu
\partial _ j \phi ^ \nu
\partial _ k \phi ^ \lambda
\xi ^ \rho \xi ^ \rho
\xi ^ \pi
\xi ^ \kappa
\cr
& + & 
3
R _ { \alpha \beta \gamma } ^ \mu
F  _ { \sigma \mu \nu \lambda}
D _ i \xi ^ \gamma
\partial _ j \phi ^ \nu
\partial _ k \phi ^ \lambda
\xi ^ \alpha
\xi ^ \beta
\xi ^ \sigma
\cr
& + &
\left.
3 R _ { \alpha \beta \gamma ; \pi } ^ \mu
\partial _ i \phi ^ \gamma
F _ { \sigma \mu \nu \lambda} 
\partial _ j \phi ^ \nu
\partial _ k \phi ^ \lambda
\xi ^ \alpha
\xi ^ \beta
\xi ^ \sigma 
\xi ^ \pi
\right \}
\end {eqnarray}

Now we wish to simply the action a bit.
We first wish to eliminate terms linear in
$\xi$.
If we add the contribution from $L _ 1 ^ { (1)}$
and $L _ 2 ^ { (1)}$, we find:

\begin {eqnarray}
L _ T ^ { (1)} &=&
{ 1 \over \alpha }
{\sqrt \gamma } \gamma ^ { ij }
g _ { \mu \nu}
\partial _ i
\phi ^ \mu
D _ j \xi ^ \nu
+
\beta \epsilon ^ { ijk }
F  _ { \sigma \mu \nu \lambda }
\partial _ i \phi ^ \mu
\partial _ j \phi ^ \nu
\partial _ k \phi ^ \lambda
\xi ^ \sigma
\cr
&=&
\partial _ i \phi ^ \mu
\left (
{{ 1 \over \alpha }
\sqrt \gamma }
\gamma ^ { ij}
g _ { \mu \nu } 
D _ j \xi ^ \nu 
+
\beta \epsilon ^ { ijk }
\partial _ j \phi ^ \nu
\partial _ k \phi ^ \lambda
\xi ^ \sigma F _ { \sigma \mu \nu \lambda}
\right)
\cr
& = &
\partial _ i \phi ^ \mu
\nabla ^ i \xi _ \mu
\end {eqnarray}
where we define $\nabla$ by:

\begin {equation}
\nabla ^ i \xi _ \mu 
= 
{ 1 \over \alpha }
{\sqrt \gamma } \gamma ^ { ij}
g _ { \mu \nu } 
D _ j \xi ^ \nu 
+
\beta
\epsilon ^ { ijk}
\partial _ j \phi ^ \nu
\partial _ k \phi ^ \lambda
\xi ^ \sigma
F _ { \sigma \mu \nu \lambda}
\end {equation}

The linear term $L ^ { (1)}$ may be set to zero, if we impose:

\begin {equation}
\nabla ^ i \partial _ i  \phi ^ \mu = 0
\end {equation}
which then defines $\phi _ { \rm cl } ^ \mu$.

Now let us add the two contributions together from the two
parts and organize them by the $\xi$ co-ordinates.

We find:

\begin {eqnarray}
L _ T ^ { (2)}
&=&
\xi ^ \alpha \xi ^ \beta L _ { \alpha \beta } ^ { (2)}
+
D _ i \xi ^ \alpha \xi ^ \beta
L _ { \alpha \beta } ^ { i (2)}
+
D _ i \xi ^ \alpha D _ j \xi ^ \beta L _ { \alpha \beta }
^ { ij (2)}
\cr
L _ T ^ { (3)}
&=&
\xi ^\alpha
\xi ^ \beta
\xi ^ \sigma
L _ { \alpha \beta \gamma } ^ { (3)}
+
D _ i \xi ^\alpha
\xi ^ \beta
\xi ^ \sigma
L _ {\alpha \beta \gamma } ^ { i (3)}
+
D _ i \xi ^\alpha
D _ j \xi ^ \beta
\xi ^ \sigma
L _ { \alpha \beta \gamma } ^ { ij (3)}
\cr
L _ T ^ { (4)}
&=&
\xi ^\alpha
\xi ^ \beta
\xi ^ \gamma
\xi ^ \delta 
L _ { \alpha \beta \gamma \delta } ^ { (4)}
+
D _ i \xi ^\alpha
\xi ^ \beta
\xi ^ \gamma
\xi ^ \delta
L _ {\alpha \beta \gamma \delta } ^ { i (4)}
\cr
&+&
D _ i \xi ^\alpha
D _ j \xi ^ \beta
\xi ^ \gamma 
\xi ^ \delta
L _ { \alpha \beta \gamma \delta } ^ { ij (4)}
+
D _ i \xi ^\alpha
D _ j \xi ^ \beta
D _ k \xi ^ \gamma 
\xi ^ \delta
L _ { \alpha \beta \gamma \delta } ^ { ijk (4)}
\end {eqnarray}

where:

\begin {eqnarray}
L _ { \alpha \beta } ^ { (2)} & = & 
{ 1 \over 2 \alpha } 
{\sqrt \gamma } \gamma ^ { ij}
R _ { \mu \alpha \beta \sigma }
\partial _ i \phi ^ \mu
\partial _ j \phi ^ \sigma 
+
{ \beta  \over 2 } 
\epsilon ^ { ijk}
D _ \alpha F _ { \beta \mu \nu \lambda}
\partial _ i \phi ^ \mu
\partial _ j \phi ^ \nu
\partial _ k \phi ^ \lambda
\cr
L _ { \alpha \beta } ^ { i (2) }
&= &
{ 3 \beta \over 2 }
\epsilon ^ { ijk}
F _ { \beta \alpha \nu \lambda}
\partial _ j \phi ^ \nu
\partial _ k \phi ^ \lambda 
\cr
L _ { \alpha \beta } ^ { ij (2) }
& = &
{ 1\over 2 \alpha }
{\sqrt \gamma}
\gamma ^ { ij}
g _ { \alpha \beta}
\end {eqnarray}

\begin {eqnarray}
L _ { \alpha \beta \gamma } ^ { (3) } & = &
{ 1 \over 6 \alpha }
{\sqrt \gamma } \gamma ^ { ij}
R _ { \mu \alpha \beta \sigma ; \gamma}
\partial _ i \phi ^ \mu
\partial _ j \phi ^ \sigma 
\cr
&+&
{ \beta  \over 6}
\epsilon ^ { ijk }
D _ \gamma D _ \beta 
F  _ { \alpha \mu \nu \lambda}
\partial _ i \phi ^ \mu
\partial _ j \phi ^ \nu
\partial _ k \phi ^ \lambda
\cr
& + &
{ \beta  \over 2 } 
\epsilon ^ { ijk }
R _ { \alpha \beta \pi } ^ \mu
F _ { \gamma \mu \nu \lambda }
\partial _ i \phi ^ \pi 
\partial _ j \phi ^ \nu
\partial _ k \phi ^ \lambda
\cr
L _ { \alpha \beta \gamma } ^ { i (3) } & = &
{ 2 \over 3 \alpha } 
{\sqrt \gamma } \gamma ^ { ij}
R _ {\mu \beta \gamma \alpha}
\partial _ j \phi ^ \mu
\cr
&+&
{ \beta } 
\epsilon ^ { ijk}
D _ { \beta } F _ { \gamma \alpha \nu \lambda}
\partial _ j \phi ^ \nu
\partial _ k \phi ^ \lambda
\cr
L _ { \alpha \beta \gamma } ^ { ij (3) } & = &
\beta \epsilon ^ {ijk}
F _ { \gamma \alpha \beta \lambda}
\partial _ k \phi ^ \lambda
\end {eqnarray}

\begin {eqnarray}
L _ { \alpha \beta \gamma \delta }
^ { (4)}
& = &
{ 1 \over 24 \alpha } 
{\sqrt \gamma }
\gamma ^ { ij }
\partial _ i \phi ^ \mu
\partial _ j \phi ^ \sigma
\left [
R _ { \mu \alpha \beta \sigma ; \gamma \delta }
+  R _ { \alpha \beta \mu } ^ \lambda
R _ { \lambda \gamma \delta \sigma }
\right ]
\cr
&+&
{ \beta  \over 4 !}
\epsilon ^ { ijk }
D _ \delta D _ \gamma D _ \alpha
F  _ { \beta \mu \nu \lambda}
\partial _ i \phi ^ \mu
\partial _ j \phi ^ \nu
\partial _ k \phi ^ \lambda
\cr
&+&
{ 9 \beta \over 4!}
\epsilon ^ { ijk }
R _ { \alpha \beta \sigma } ^ \mu
D _ \gamma F _ { \delta \mu \nu \lambda }
\partial _ i \phi ^ \mu
\partial _ j \phi ^ \nu
\partial _ k \phi ^ \lambda
\cr 
&+&
{3 \beta \over 4!}
\epsilon ^ { ijk}
R _ { \alpha \beta \sigma ; \delta }
^ \mu
\partial _ i \phi ^ \sigma 
F _ { \gamma \mu \nu \lambda}
\partial _ j \phi ^ \nu
\partial _ k \phi ^ \lambda
\cr
L _ { \alpha \beta \gamma \delta } ^ { i(4)}
& = &
{ 1 \over 4 \alpha } 
{\sqrt \gamma } \gamma ^ { ij}
R _ { \mu \alpha \beta \gamma ; \delta }
\partial _ i \phi ^ \mu
+
{18 \beta \over 4!}
\epsilon ^  { ijk }
R _ {\beta \delta \pi } ^ \mu
F  _ { \delta \mu \alpha \lambda}
\partial _ j \phi ^ \pi 
\partial _ k \phi ^ \lambda
\cr
&+&
{ 9 \beta  \over 4!}
\epsilon ^ { ijk}
D _ \delta D _ \beta F _ { \gamma \alpha \nu \lambda }
\partial _ j \phi ^ \nu
\partial _ k \phi ^ \lambda
+
{ 3 \beta \over 4!}
\epsilon ^ { ijk}
R _ { \beta \gamma \alpha } ^ \mu
F  _ { \delta \mu \nu \lambda}
\partial _ j \phi ^ \nu
\partial _ k \phi ^ \lambda
\cr
L _ { \alpha \beta \gamma \delta } ^ { ij(4)}
&=&
{ 1 \over 6 \alpha }
{\sqrt \gamma } \gamma ^ { ij }
R _ { \alpha \beta \gamma \delta } 
+
{ 18 \beta  \over 4!}
\epsilon ^ { ijk}
D _ \delta F _ { \gamma \alpha \beta \lambda}
\partial _ k \phi ^\lambda
\cr
L _ { \alpha \beta \gamma \delta } ^ { ijk(4)}
&=&
{ 6 \beta \over 4!}
\epsilon ^ { ijk }
F  _ { \delta \alpha \beta \gamma }
\end {eqnarray}

Before we can begin to set up the  
perturbation series, we must first diagonalize the
quadratic term $L _ T ^ { (2)}$.
The space-time matrix $g _ { \mu \nu}$ can be eliminated
in favor of the usual vierbein $ e _ { \mu } ^ a$,
where the Roman index $a$ represents tangent space indices.

Now let us replace $g _ { \mu \nu}$
with $e _ \mu ^ a e _ \nu ^ a$.
With a little bit of algebra, we can move the
vierbein past the derivative and prove the identity:

\begin {eqnarray}
e _ \mu ^ a D _ i \xi ^ \mu 
&=& e _ \mu ^ a \left ( \partial _ i \xi ^ \mu 
+ \partial _ i \phi ^ \lambda 
\Gamma _ { \lambda \nu } ^ \mu \xi ^  \nu \right) 
\cr
& = &
\partial _ i \xi ^ a
+ 
\partial _ i \phi ^ \mu 
\omega _ \mu ^ { ab } 
\xi ^ a
= D _ i \xi ^ a
\end {eqnarray}
where $\xi ^ a = e _ \mu ^ a \xi ^\mu$ and where:

\begin {equation}
\omega _ \mu ^ { ab } 
= - (\partial _ \mu e _\nu ^ a )
e ^ { b \nu }
+
e _ \nu ^ a \Gamma _ { \mu \beta } ^ \nu 
e ^ { b \beta}
\end {equation}
which is self-consistent with the equation
$D _ \mu e _ \nu ^ a = 0$, as desired.

In this fashion, we can now write everything with tangent
space indices. We find that the only change is that
we must replace Green letters $\alpha, \beta,\gamma , \delta$ with
Roman letters $a,b,c,d$.

For the general case, we find:

\begin {equation}
L _ T ^ { (n)} =
\sum _ {k=0} ^ {n-1}
D _ { i _ 1 } \xi ^ { a _ 1 }
D _ { i _ 2 } \xi ^ { a _ 2 } 
\cdots
D _ { i _ k} \xi ^ { a _ k }
\cdots
\xi ^ { a _ n}
L _ { a _ 1 \cdots a _ n } ^ 
{ i _ 1 i _ 2 \cdots i _ k (n) }
\end {equation}
where $a _ i$ are defined in the tangent space.

\section {Regularization}
Now that we have power expanded the original action
in terms of $\xi ^ a$, where $a$ is the tangent space index
on curved space-time, 
we must now integrate
over the quantum field $\xi ^ a$ defined on the tangent space, 
which will leave us
with divergent terms whose structure we wish to
analyze.

We will power expand around the term:
\begin {equation}
L _ 1 ^ { (2) } =
{ 1 \over 2\alpha } 
{\sqrt \gamma }
\gamma ^ { ij }
D _ i \xi ^ a
D _ j \xi  ^ a
\end {equation}
where the space-time metric $g _ { \mu \nu}$ has been absorbed into
the vierbeins.

If we perform the integration over $\xi ^ a$, then 
(subject to a regularization scheme):

\begin {equation}
\langle T \xi ^ a ( x ) \xi ^ b ( x ' ) \rangle
\sim \alpha \Lambda \delta ^ { ab} + ...
\end {equation}
where the right hand side is linearly divergent via some
large momentum scale $\Lambda$, and there are
important corrections to this equation
crucially dependent on the regularization scheme.

Then the first term
in $L _ T ^ { (2)}$ contributes the following
term:
\begin {equation}
L _ { ab } ^ { (2) } 
\langle T \xi ^ a \xi ^ b \rangle
\end {equation}
which in turn yields the two equations:

\begin {equation}
\langle T \xi ^ a \xi ^ b \rangle
{ 1 \over \alpha }
{\sqrt \gamma }
\gamma ^ { ij }
R _ { \mu ab \sigma} \partial _ i \phi ^ \mu
\partial _ j \phi ^ \sigma
\end {equation}
and:
\begin {equation}
\beta
\langle T \xi ^ a \xi ^ b \rangle
\epsilon ^ { ijk }
\partial _ i \phi ^ \mu 
\partial _ j \phi ^ \nu 
\partial _ k \phi ^ \lambda 
D _ a F _ { b \mu \nu \lambda }
\end {equation}

Unlike the superstring, there is no conformal symmetry
by which we can set this divergent term to zero.
In this paper, we will simply set the lowest order
divergent term to zero by fiat. This is a weakness in this
approach. This, in some sense, 
defines the model, i.e. the theory can only
propagate on certain background fields which set the
lowest order divergent term to zero.

However, for the supermembrane, we will show in a later paper
that there is enough supersymmetry to allow us
to set this divergent term to zero, so we have:

\begin {equation}
R _ { \mu \nu } = 0 ; \quad
D ^ a F _ { abcd} = 0
\end {equation}

The second equation is just the equation of motion for the
anti-symmetric field, as expected. However, the first equation
is rather troubling, since
there should be a term proportional to $F ^ 2$.
The fact that this term is missing means that the equations of motion are
actually inconsistent.
There exists no action involving $g _ { \mu \nu}$ and
$F  _ { \alpha \beta \gamma \delta}$ which
yields these equations of motion.
Thus, we must carefully analyze our regularization
scheme and go to higher interactions.
This is different from the superstring case, where
the one-loop results are sufficient to yield
self-consistent equations of motion.
For the membrane, we find that we must go to
two loops in order to obtain self-consistent results.

Now let us generalize this result to higher orders by
carefully introducing a regularization scheme.
There is a problem with dimensional regularization, however.
If we analytically continue the integral:

\begin {equation}
\int { d ^ d p 
\over
( p ^ 2 + m ^ 2 ) }
\sim
\Gamma ( 1 - { d \over 2 } )
\end {equation}
we find that it is finite for $d = 3$.
It diverges with a pole at $d =4$, but is formally
finite for odd dimensions.
This strange result does not change for higher loops, since
multiple integrals over the momenta yield
factors of $\Gamma ( k )$,
where $k$ is half-integral, which is again finite for
$d=3$.
Furthermore, when we introduce supersymmetry, we find that
dimensional regularization does not respect this symmetry,
which only holds at $d =3$ for supermembranes. Hence,
dimensional regularization poses some problems.
In fact, supersymmetry is so stringent, it appears that finding
a suitable regularization method is problematic for any method.

We will, instead, use standard point-splitting and proper time methods,
separating the points on the world volume at which the
various $\xi ^ \mu ( x )$ meet at a vertex.
This, of course, will violate general covariance
and supersymmetry by point-splitting.
However, point-splitting methods are convenient since
the divergence within a Feynman integral occur when
fields are defined at the same world volume
point, i.e. $x \rightarrow y$ on the world volume.

Then the two-point Green's function  can be written as:
\begin {equation}
\langle
T \xi ^ a ( x ) 
\xi ^ b ( x ' ) \rangle =
- i G ^ { ab } 
( x - x ' )
\end {equation}
where $x$ and $x'$ represents points on the three dimensional
world volume, and:

\begin {equation}
{ 1 \over \alpha }
\left [
- D _ i ^ { ab } 
{\sqrt \gamma }
\gamma ^ { ij }
D _ j ^ { bc } 
+
{\sqrt \gamma }
\left ( {\zeta } R
+  m ^ 2 \right )
\right ]
G ^ { ac } ( x , x ' )
=
\delta ^ 3 ( x - x ' ) 
\delta ^ { ab } 
\end {equation}
(Although the theory is massless, notice 
that we added in a small mass $m ^2$ in order to
handle infrared divergences. In non-linear sigma
models of this type, it can be shown that this
mass regulartor cancels
against other terms in the perturbation theory.
Notice that we introduce a parameter
$\zeta$ which takes into account the
curvature on the
world volume. This term will
be of interest when we introduce fermions
into our formalism. However, here we can set this term $\zeta$ to zero
for our case.)

The solution of this Green's function is complicated
by two facts.
First, the Green's function is defined over both a curved
three dimensional world volume manifold and 11 dimensional space-time
manifold, and hence we have to
use the formalism developed for general relativity.

Second,
the Green's function will in general introduce
unwanted curvatures on the world volume.
This is because the covariant derivative $D _ i$
contains the connection field $\partial _ i \phi ^ \mu
\omega _  { \mu } ^ { ab }$.
If we set this equal to $A _ i ^ { ab }$,
then $D _ i ^ { ab  } = \partial _ i \delta ^ { ab } +
A _ i ^ { ab }$, which is the familiar covariant
derivative in $O(D)$ gauge theory.
Thus, when we invert 
the opertor $D _ i {\sqrt \gamma } \gamma ^ { ij } D _ j$,
we encounter gauge invariant
terms like
the square of 
$R _ { ij } ^ { ab } $, where
\begin {equation}
[ D _ i , D _ j ] ^ { ab }
= R _ { ij } ^ { ab }
\end {equation}

In two dimensions, curvature terms of this sort do not contribute
to the perturbation expansion to lowest order when $d=2$ [8].
However, these terms do in fact contribute to 
the perturbation expansion when $d = 4$ [10]. In fact, the
presence of these terms renders the quantum theory of
the $d=4$ non-linear sigma model non-renormalizable, since
they introduce new counter-terms not present in the original
action.

Given the potential problems with this term, let us
introduce the proper time formalism [11]. Let $s$ be the
Schwinger proper time
variable:
\begin {equation}
\tilde G ^ { ab }
( x , x ' )
=
( \gamma (x) ) ^ { 1/4}
G ^ { ab } ( x , x ' )
( \gamma ( x ' ) ) ^ { 1/4 } 
= 
i \alpha \int _ 0 ^ \infty
ds
\langle x , s |
x ' , 0 \rangle ^ { ab }
\end {equation}
where $\gamma = {\rm det } \gamma ^ { ij }$
and we choose a positive metric on the world volume.

We impose the boundary condition:
\begin {equation}
\langle x , 0 | x  ' , 0 \rangle ^ { ab } 
= \delta ^ { ab } \delta ^ 3 ( x - x ')
\end {equation}

Now let us assume the ansatz for the
Green's function:

\begin {eqnarray}
\langle x, s |
x ' , 0 \rangle 
&=&
{ i \over 
(4 \pi i s ) ^ { d/2} }
\gamma (x ) ^ { 1/4}
\Delta ^ { 1/2} ( x , x ')
\gamma ( x ' ) ^ { 1/4}
F ^ {ab}
( x, x'; is )
\cr
&\times&
{\rm exp }
\left ( - { \sigma ( x , x ' ) \over 2 i s }
- is m ^ 2 \right)
\end {eqnarray}
where $\sigma ( x, x ' ) $ is one-half the square of the
distance along the geodesic between $x$ and $x'$,
where $\Delta$
is given by:

\begin {equation}
\gamma ( x ) ^ { 1/4}
\Delta ( x , x ' ) 
\gamma ( x ' ) ^ { 1/4}
=
-
{\rm det }
\,
\left ( 
-\sigma _ { , i,j } ( x , x ' ) \right )
\end {equation}
and:
\begin {equation}
F ( x , x ' ) ^ { ab } = \sum _ { n=0} ^ \infty
( is ) ^ n a _ n ^ { ab } ( x , x ' )
\end {equation}

The object of this section is to power expand
the Green's function in terms of 
$\sigma ( x , x ')$. In particular,
$\sigma$ obeys a number of useful identities, among them:
\begin {equation}
\sigma _ { i }  \sigma ^ { i} = 2 \sigma
\end {equation}
where $\sigma _ i = \partial _ i \sigma$, and we
raise and lower indices via $\gamma ^ { ij }$. From this
identity, we can establish a large number of identities
for various derivatives of $\sigma$.

When calculating Feynman diagrams, we will find that they
diverge according to inverse powers of $\sigma$.
Hence, we can compare the large momentum cut-off 
$\Lambda$ to $\sigma$, i.e. 

\begin {equation}
\Lambda \sim
\sigma ^ { - 1/2}
\end {equation}

This new Green's function satisfies the \lq\lq Schrodinger"
equation:

\begin {equation}
- {\partial \over \partial is }
\langle x,s | x ' , 0 \rangle =
H \langle x , s | x ',  0 \rangle
\end {equation}
where the \lq\lq Hamiltonian" is given by:

\begin {equation}
H = - \gamma ^  { -1/4}
D _ i \gamma ^ { 1/2} \gamma ^ { ij }
D _ j \gamma ^ { - 1/4 } 
+ 
{\zeta } R + m ^ 2
\end {equation}

If we insert the expansion of the Green's function into the
defining equation for the Green's function, we are left with
a constraint on the undetermined function $F$:

\begin {equation}
- { \partial F \over \partial is }
=
\zeta R F + { 1 \over is }
\sigma ^ { , i} F _ {, i }
- { 1 \over \Delta ^ { 1/2} }
\left (\Delta ^ { 1/2} F \right ) _ { , i} ^ {; i }
\end {equation}

Inserting the power expansion for $F$ into this expression, we now have
a recursive relation among the $a _ n$ coefficients appearing within
$F$:

\begin {equation}
\sigma ^ { i }
a _ { n+1 , i }
+ (n+1) a _ { n+1}
=
{ 1 \over \Delta ^ { 1/2}}
\left ( \Delta ^ { 1/2} \right ) _ { i  } ^ { ; i  }
- { \zeta } R a _ n
\end {equation}

Now let us solve the $a _ n$ iteratively.
The equation for $a _ 0 ^ { ab } $ give us:

\begin {equation}
\sigma ^i D _ i a _ { 0 } = 0
\end {equation}

The goal of this exercise is to extract out the
divergent terms within the Green's function.
Let us, therefore, slowly let $x$ 
approach $x '$.
Repeated differentiation of the previous formula yields:

\begin {eqnarray}
{\rm lim } _ { x \rightarrow x ' }
\,\, D _ i a _ { 0 } &=& 0
\cr
{\rm lim } _ { x \rightarrow x ' }
\,\, D _ i \gamma ^ { ij } D _ j a _ { 0 }
&=& - { 1 \over 2 } 
R _ { ij } ^ { ab } 
\cr
{\rm lim } _ { x \rightarrow x ' }
\,\, { { { a _ { 0 , ; i } } ^ { i } } _ { ; j } } ^ j 
&=& { 1 \over 2 } \, {\rm Tr } \, R _ { ij } R ^ { ij } 
\end {eqnarray}
where $R _ { ij } ^ { ab }
= [ D _ i , D _ j ] ^ { ab } $
and:

\begin {eqnarray}
\sigma & \rightarrow & { 1 \over 2 } ( x - x ' ) ^ 2
\cr
\sigma _ { i ; j } &\rightarrow& \gamma _ { ij }
\cr
\sigma _ { i ; j ; k ; l } &\rightarrow&
{ 1 \over 2 } \left (
R _ { lijk} + R _ { kijl} \right)
\cr
\Delta & \rightarrow & 1
\end {eqnarray}

After a certain amount of algebra, we find the desired result
for the coefficients $a _ n ^ { ab }$:

\begin {eqnarray}
a _ 0 & \rightarrow & \delta ^ { ab }
\cr
a _ 1 & \rightarrow & { 1 \over 6 } R 
- \zeta R
\cr
a _ 2 &\rightarrow &
{ 1 \over 2 } \left[ ( { 1 \over 6 } -   \zeta ) R \right ]  ^ 2
+ { 1 \over 6 }
( { 1 \over 5 } - \zeta ) R _ { i } ^ { ; i}
+
{ 1 \over 12 }
\, {\rm Tr } \, R _ { ij } R ^ { ij }
\cr
& - &
{ 1 \over 180}
R _ { ij } R ^ { ij }
+ { 1 \over 180 }
R _{ ijkl} R ^ { ijkl}
\end {eqnarray}

Now we wish to insert these values for $a _ n ^ { ab}$ into
the expression for the Green's function, in order to
see how badly it diverges as a function of
$\sigma$.
In the limit as $x \rightarrow x ' $, many 
terms drop out, and we are left with:

\begin {eqnarray}
\tilde G ( x , x ' ) ^ { ab } 
&=&
i \alpha \int _ 0 ^ { \infty}
ds \,
\langle x , s | x ; 0 \rangle ^ { ab } 
\cr
&=&
i  \alpha \int _ 0 ^ \infty
ds \,
{ i {\sqrt \gamma } \over 
( 4 \pi i s ) ^ { d/2} } 
{\rm exp } \,
\left ( - { \sigma \over 2 i s } - is m ^ 2 \right)
\sum _ { n=0} ^ \infty (is) ^ n  a _ n ^ { ab }
\end {eqnarray}

We now use the integral:
\begin {equation}
\int _ 0 ^ \infty
dx \,
x ^ { \nu - 1 } 
{\rm exp } 
\,
\left (
- { i \mu \over 2 } [ x + (\beta ^ 2 / x ) ] \right)
=
- i \pi \beta ^ \nu e ^ { i \nu \pi / 2 }
H _ { - \nu } ^ { (2) } ( \beta \mu )
\end {equation}
where $H ^ { (2)}$ is a Bessel function of the third kind,
or a Hankel function.

For the case of interest, $d =3$, we have:

\begin {equation}
\tilde G ( x, x ' ) ^ { ab  } 
=
\alpha { \pi {\sqrt \gamma} \over ( 4 \pi ) ^ { 3/2} }
\sum _ { n = 0} ^ \infty
a _ n ^ { ab } 
\left ( { - \sigma \over 2 m ^ 2 } \right )
^ { (1/2)(n - 1/2) }
H _ { -n  + 1/2 } ^ { (2)} 
\left (
{\sqrt { - 2 m ^ 2 \sigma } } \right)
\end {equation}

We now make the definitions $\mu = 2 m ^ 2$,
$\beta = ( - \sigma / 2 m ^ 2 ) ^ { 1/2} $,
$\nu = n + 1 - (d/2)$.
To find the power expansion in $\sigma$, we use the fact that:
\begin {equation}
H _ { - n + 1/2 } ^ { (2) } =
- i ( -1 ) ^ { n - 1 } \left (
{ 2 z \over \pi } \right ) ^ { 1/2 } 
\left ( j _ { n-1}  - i y _ { n-1 } \right)
\end {equation}
(where we set  $ z = {\sqrt { - 2 m ^ 2 \sigma }}$)
and the fact that:

\begin {eqnarray}
j _ { n } (z ) &=&
z ^ n \left ( - { 1 \over z } { d \over dz } \right ) ^ n 
{  {\rm sin} \, z   \over z }
\cr
y _ { n } (z) &=&
- z ^ n \left ( - { 1 \over z } { d \over dz }
\right ) ^ n 
{  {\rm cos } \, z   \over z }
\end {eqnarray}

In particular, we find that the propagator in curved space
is linearly divergent in momentum, as expected.
The troublesome terms, $a _1$ and $a _ 2$, we see, are 
finite in three
dimensions, and so can be dropped from our discussion
(which is not the case for $d =4$). Although they
ruin the renormalization program for four dimensions,
we find that they drop out in three dimensions.

For completeness, we present the entire series:

\begin {eqnarray}
\tilde G ( x , x ' ) ^ { ab } 
& = &
\alpha { i { \sqrt { 2 \pi \gamma } \over ( 4 \pi ) ^ { 3/2 } }}
\left \{  { e ^ { -i z } \over \sqrt \sigma }
a _ 0 ^ { ab } \right.
\cr
& + &
\left .
\sum _ { n=1}
^ \infty
( -1) ^ n a _ n ^ { ab } 
\left ( { z \over - 2 m ^ 2 } \right ) ^ { 2 n - 1 } 
\left ( - { 1 \over z } { d \over dz } \right ) ^ { n-1}
{ e ^ { - iz } \over z } 
\right \}
\end {eqnarray}

Notice that, as $x \rightarrow x ' $, we find that
the integral diverges linearly with the momentum,
but the troublesome curvature term involving
$R _ { ij } ^ { ab}$ do not contribute (as they do in
four dimensions).
Hence, from now on, we can simply use the fact that the
propagator diverges linearly with momentum.

Lastly, we can also use this formalism to compute
two point functions involving derivatives.
If we have two point functions like:

\begin {equation}
\langle T \partial _ i \xi ^ a (x) \partial _ j \xi ^ b (y)
\rangle
= \partial _ i ^ x \partial _ j ^ y \langle T  \xi ^ a (x)
\xi ^ b ( y) \rangle = \partial _ i ^ x \partial _ j ^ y
\tilde G ^ { ab } ( \sigma)
\end {equation}
where $\tilde G ^ { ab } ( \sigma)$ is the propagator,
then we can, for small distance separations, use the fact that
$\sigma ( x,y) \sim (1/2) ( x -y ) ^ i \gamma _ { ij }
( x - y) ^ j$, so that:
\begin {equation}
\partial _ i ^ x \partial _ j ^ y \sigma ( x,y)
\sim
- \gamma _ { ij }
\end {equation}
By taking repeated derivatives of the propagator
as a function of the separation $\sigma$, one
can therefore contruct the contraction of an arbitrary
number of $\xi$ fields.

\section {Two Loop Order}

We saw earlier that the one loop result was inconsistent.
An action of the form $R + F _ { \mu \nu \alpha \beta} ^ 2$
cannot have equations of motion given by 
$R _ { \mu \nu} = 0$ and $D ^ \mu F _ { \mu \nu \alpha \beta} = 0$
We must therefore probe the two loop result to see if we can
re-establish the consistency of the model.

Consider first the case of two external lines $N=2$.

The contraction of the term $L _ { 2 , \alpha \beta} ^ { (3)}$
yields:
\begin {equation}
L _ { 2 , a a  } ^ { i (3)}  \sim  
\epsilon ^ { ijk }
F _ { a a \nu \lambda  } 
\partial _ j \phi ^ \nu
\partial _ k \phi ^ \lambda
= 0
\end {equation}
which vanishes by the anti-symmetry of the $F$ tensor.

The most interesting two loop graph is given by the
contraction of 
$ L _ { 2 \alpha \beta \gamma } ^ { ij (3) } $
with itself. This gives us the contraction:

\begin {equation}
\epsilon ^ { ijk }
F _ { \gamma \alpha \beta \lambda } 
\partial _ k \phi ^ \lambda
\langle
T D _ i \xi ^ \alpha (x)
D _ j \xi ^ \beta (x)
\xi ^ \gamma (x) 
D _ { \bar i } \xi ^ { \bar \alpha} (y)
D _ {\bar j } \xi ^ {\bar \beta} (y)
\xi ^ {\bar \gamma} (y)
\rangle
\epsilon ^ { \bar i \bar j  \bar k  }
F _ { \bar \gamma \bar \alpha \bar \beta \bar \lambda } 
\partial _ {\bar k}  \phi ^ {\bar \lambda}
\end {equation}
when $x \rightarrow y$.

If we perform the contractions over $\xi$, we find, with a little
bit of work, the following
result:
\begin {equation}
\beta ^ 2 \alpha ^ 3
\Lambda ^ 4 {\sqrt \gamma } \gamma ^ { ij }
\partial _ i \phi ^ \lambda
\partial _ j \phi ^ {\bar \lambda}
F _ { cab \lambda}
F _ { \bar \lambda}  ^ { cab }
\end {equation}
Notice that the divergence can be absorbed into a
rescaling: $\alpha = \alpha _ R / \Lambda $.
(In the next section, we will see that the leading
divergences can in fact be absorbed by this
rescaling to all orders.)

After rescaling, we find that the equation of motion of the
graviton is given by:
\begin {equation}
R _ { \mu \nu } + { 1 \over 4 } 
F _ { \mu \alpha \beta \gamma }
F _ \nu ^ { \alpha \beta \gamma} = 0
\end {equation}
(Unfortunately, the term
proportional to $g _ { \mu \nu}$ does not appear
in the equations of motions, signalling a possible
inconsistency. This is normally solved for the superstring case
by adding an another field, the dilaton. 
We will see that this possible inconsistency vanishes
for the supermembrane case.)

There is also a self-consistency between the equations of motion for
the metric and the anti-symmetric field
which must be re-established at every loop order,
and hence this provides a powerful check on the correctness of any
model of membranes.

Higher order graphs are easy to construct but more tedious
to evaluate. We will  present the contractions necessary
to perform two and three loop calculations, but will
not explicitly compute the graphs.

For example, the $R^2$ and $DR$ two loop terms are contained in
the contraction of $L _  { 1 , \alpha \beta \gamma \delta } ^ { (4)}$, so we
have the two loop contribution:

\begin {equation}
{ 1 \over 24 \alpha } 
{\sqrt \gamma }
\gamma ^ { ij }
\partial _ i \phi ^ \mu
\partial _ j \phi ^ \sigma
\left [
R _ { \mu \alpha \beta \sigma ; \gamma \delta }
+  R _ { \alpha \beta \mu } ^ \lambda
R _ { \lambda \gamma \delta \sigma }
\right ]
\langle
T \xi ^ \alpha \xi ^ \beta 
\xi ^ \gamma \xi ^ \delta \rangle
\end {equation}

Two loop curvature terms are also contained in the
contraction of the square of
$L _ { 1 \alpha \beta \gamma} ^ { ( 3) }$.
This contraction yields:

\begin {eqnarray}
\langle L _ 1 ^ { (3) } L _ 1 ^ { (3) } \rangle
& = & 
\left ( { 2 \over 3 \alpha } \right ) ^ 2
{\sqrt \gamma } \gamma ^ { ij}
R _ {\mu \beta \gamma \alpha}
\partial _ j \phi ^ \mu
\langle T D _ i \xi ^ \alpha (x)
\xi ^ \beta (x)
\xi ^ \gamma (x) 
\cr
& \times &
D _ {\bar i}  \xi ^ {\bar \alpha} (y)
\xi ^ {\bar \beta} (y)
\xi ^ {\bar \gamma} (y)
\rangle
{\sqrt \gamma } \gamma ^ { \bar i \bar j}
R _ { \bar \mu \bar \beta \bar \gamma \bar \alpha}
\partial _ {\bar j } \phi ^ {\bar \mu}
D _ {\bar i}  \xi ^ {\bar \alpha}
\xi ^ {\bar \beta}
\xi ^ {\bar \gamma}
+
...
\end {eqnarray}

These two terms give us 
two loop correction terms to the curvature tensor,
yielding complicated combinations of
$R ^2$ and $DR$ terms.

Lastly, we can also calculate the two and three loop contribution
for the anti-symmetric field. 
For example, two loop corrections 
to the equations of motion are given by contracting
$L _ { 2 \alpha \beta \gamma \delta } ^ { (4)}$
with two propagators.
This term is contained within:

\begin {eqnarray}
\langle L _ { 2 } ^ { (4)} \rangle 
& =&
\beta \epsilon ^ { ijk }
\left \{ 
{ 1 \over 4 !}
D _ \delta D _ \gamma D _ \alpha
F  _ { \beta \mu \nu \lambda}
\partial _ i \phi ^ \mu
\partial _ j \phi ^ \nu
\partial _ k \phi ^ \lambda
\right.
\cr
&+&
\left.
{ 9 \over 4!}
R _ { \alpha \beta \sigma } ^ \mu
D _ \gamma F _ { \delta \mu \nu \lambda }
\partial _ i \phi ^ \mu
\partial _ j \phi ^ \nu
\partial _ k \phi ^ \lambda
\right \}
\langle T \xi ^ \alpha 
\xi ^ \beta
\xi ^ \gamma
\xi ^\delta
\rangle
+ ...
\end {eqnarray}
This gives us terms like $RF$ and $DDDF$.

Similarly, we can also contract
over the square of
$L _ { 2 \alpha \beta \gamma \delta } ^ { ij (4)}$,
which will give us a $FDDF$ term. It is contained within:

\begin {eqnarray}
\langle ( L _ { 2 \alpha \beta \gamma } ^ { (4) } ) ^ 2 \rangle
&=&
\left ( { 18 \beta \over 4!}\right )^ 2 
\epsilon ^ { ijk}
D _ \delta F _ { \gamma \alpha \beta \lambda}
\partial _ k \phi ^\lambda
\langle T D _ i \xi ^ \alpha (x)
D _ j \xi ^\beta (x)
\xi ^ \gamma (x)
\xi ^ \delta (x) 
\cr
& \times & 
D _ {\bar i} \xi ^ {\bar \alpha} (y)
D _ {\bar j} \xi ^ {\bar \beta} (y)
\xi ^ {\bar \gamma} (y)
\xi ^{\bar  \delta} (y)
\rangle
\epsilon ^ { \bar i \bar j \bar k}
D _ {\bar \delta} F _ { \bar \gamma \bar \alpha \bar \beta \bar \lambda}
\partial _ {\bar k} \phi ^ {\bar \lambda}
+ ... 
\end {eqnarray}

\section {Power Counting}

Now let us analyze the divergence of graphs to all orders in
perturbation theory. Because the coupling constant has
negative dimension, we can always increase the
degree of divergence of any multi-loop graph by
adding more insertions. In this sense, the theory is
not renormalizable. But we will see in this section
how many divergences we can absorb via the
coupling constant $\alpha$ and $\beta$.

Consider first the Lagrangian $L _ 1$
with only the metric tensor, without the anti-symmetric
field.
Let $L$ be the number of 
loops in an arbitrarily complicated Feynman graph.
Then its contribution to the over-all divergence is
$3L$, due to $d ^ 3 p$.
Let $I$ be the number of internal lines in the graph.
So its contribution is $-2L$ due to $1/ p ^ 2$. 
Let $V _ n$ be the number of $n$-point vertices in the
graph. Since each $n$-point graph in the action
has two momenta
associated with it, it can contribute at most
$2 V _ n$.
Let $E$ equal the number of external lines in the graph.
Since each external line 
subtracts off a line
which could have become an internal line, it contributes
$- N$.
Then the superficial divergence of any graph $D$
is given by:

\begin {equation}
D = 3 L - 2 I + 2 \sum _ { n=3} ^ \infty V _ n
- N
\end {equation}

Now calculate the number of momentum integrations.	
Each internal line contributed $d ^ 3 p$.
Each $n$-point vertex contributes a 
momentum-conserving delta function $\delta ^ 3 ( \sum p _ i)$,
which deletes three momentum integrations per vertex.
And then there is one over-all conservation of momentum factor.
The sum of these integrations, in turn, contributes an over-all
$ (d ^ 3 p _ i ) ^ L $
momentum integration for the loops. Thus, we have:

\begin {equation}
L = I - \sum _ { n=3} ^ \infty V _ n + 1
\end {equation}

Now insert the second equation into the first, and we obtain:

\begin {equation}
D = L - N + 2
\end {equation}
Notice that the degree of divergence $D$ is just a function
of the number of loops $L$ and the number of external
lines $E$.

Now let us see if we can re-absorb this divergence
into the coupling constant $\alpha$.
Let $\Lambda$ be the momentum cut-off for the graph.
Recall that the perturbation expansion parameter is
$\alpha$.
Then the leading divergences of the $N$-point
amplitude $A _ N$,
symbolically speaking, diverge as:

\begin {equation}
A _ N = \sum _ { L = 1} ^ \infty
\alpha ^ { L-1} A _ {N,L}
\end {equation}
where we only compute the loop corrections.

We have just shown that $A _ {N,L}$
diverges as:
\begin {equation}
A _ {N,L} \sim \Lambda ^ { L-N + 2} \tilde A _ { N,L}
\end {equation}

Now let us re-define the coupling constant as:
\begin {equation}
\alpha \sim { \alpha _ R \over \Lambda}
\end {equation}
which we performed in the last section for the single loop.

Rescaling the 
graph, we now have:

\begin {equation}
A _ N = \Lambda ^ { 3-N}
\left ( \sum _ { L=1}
\alpha ^ { L-1}
\tilde A _ { N,L} \right)
\end {equation}

Thus, the leading superficial divergence can be absorbed into
$\alpha$ by a rescaling.
The larger $N$, the faster the graph converges.
In particular, we see that the amplitude is formally finite
for $N = 3$ and beyond, but diverges still for $N=2$.
We can eliminate the $N=2$ divergence by simply declaring that
the background fields obey the standard equations of motion,
thereby defining the model.

Now let us generalize the simple power counting to the
general case, including the anti-symmetric field.
The power counting is much worse, since we now have
3 momenta attached to each vertex function, rather than 2.
The leading divergences all come from this sector.

So the degree of divergence is now given by:
\begin {equation}
D = 3L - 2 I + 3 \sum _ { n=3} ^ \infty
 V _ n + X _ N
\end {equation}
(If some of the lines on the vertex are external lines,
this reduces the degree of divergence of the graph,
so we have to compensate this by adding in $X _ N$.
For example, $X _ 2 = -2, X _ 3 = -3 $.)

The number of momentum integrations is given by:
\begin {equation}
L = I - \sum _ { n=3} ^ \infty
V _ n +1
\end {equation}

Notice that we can no longer cancel both $I$ and $\sum V _n$
to arrive at a simple relationship involving just $L$ and
$N$.
Thus, we need one more constraint to eliminate the
vertex factors.

Let us count the number of lines in a graph.
Each of the $V _ n$ vertices contributes $n$ lines to the graph.
Thus, they collectively contribute $\sum n V _ n$ lines
to the graph. When two of these vertices $V _n$ are joined,
they form an internal line, which is therefore counted
twice.
This means that the sum $\sum n V_n$ counts each
internal line twice, and each external line once (since
external lines are not paired off).
Thus, we have:

\begin {equation}
\sum _ { n=3} ^ \infty
n V _ n = 2 I + N
\end {equation}

By examining these three sets of equations, we see that, in 
general, it is not possible to eliminate all the
$V _ n$ in  a graph.
Therefore, we will only concentrate on the leading divergence
within a graph and ignore lower order divergences.

Let us see which vertices contribute the most to the over-all
divergence.
A vertex $V _n$ contains 3 momenta.
Let us say that we replace it with two small
vertices $V _ {n_1}$ and $V _ { n+2}$ which are joined
by an internal line. 
The over-all contribution from these two attached vertices
is given by $ 3 + 3 - 2 = 4$, where the $-2$ comes from
$1/p ^2$.
Thus, we can always increase the over-all divergence of a graph
by replacing $V _ n$ with pairs of smaller $n$-point vertices. 
This process can be continued, until we are left with a graph with
only $V _4 $ and $V _5$ vertices left.
Thus, the leading divergence is now given with only
$V _4$ and $V _5$.
If we eliminate $V _4$, we are left with:

\begin {equation}
D = 2L - { 1\over 2 } V _ 5 + { N \over 2 }
+ X _ N + 1
\end {equation}
(Notice that this equation depends on whether the overall
number of vertex lines is even or odd. If it is even,
then $V _ 5 = 0$.)

In this way, we can compute the over-all divergence of a graph.
However, there is simple short-cut we can use.
If we examine the perturbation expansion of $L _ 1$ and
$L _ 2$, we see that the primary difference is that
the internal vertices of $L _1 $ contain two derivatives,
while the internal vertices of $ L _ 2$ contain three derivatives.
Because the coupling
constant $ 1/ \alpha$ appears in front of each term in 
$L _ 1$, we see that each internal vertex function
diverges, at most, like $\Lambda ^ 3$.
But if we let $\beta$ remain a finite constant, we see
that each internal vertex in $L _ 2$ diverges
as $\Lambda ^ 3$ as well.
Thus, by only rescaling $\alpha$ but keeping $\beta $ finite,
we see that the divergence of the purely metric theory is
identical to the theory coupled to anti-symmetric tensor fields.
($\beta$, although it is finite, will ultimately be fixed by
requiring consistency in the equations of motion of the
background fields).

\section {Conclusion}

In field theory, the study of non-renormalizable Lagrangians,
such as the four-fermion model, or massive vector 
theories, has given us insight in deep physical processes.
Likewise, the bosonic membrane action, by naive power counting,
is non-renormalizable on the world volume, but
may give us insight into M-theory.
Although the bosonic membrane theory is ultimately probably
not a consistent quantum theory, the techniques we have used here
will generalize to the supermembrane case.

In this paper, we have expanded the bosonic membrane action
around Riemann normal co-ordinates, treating the theory as
a non-linear sigma model, and calculated the
regularized propagator and higher loop graphs.

In particular, we found:

a) The standard dimensional regularization method 
apparently breaks down at $d = 3$, where the
Gamma function no longer has a pole. Instead, we 
developed the proper time and point-splitting formalism
in curved space for the $d=3$ membrane action. 
Although we lost general covariance, this gave us
an intuitive way in which to isolate all the divergences
of higher graphs, since the singularities emerge when
two fields touch on the world volume. It is then a simple matter
to analyze complicated graphs visually
and isolate their divergences.

b) The renormalization program in four dimensions 
for the non-linear sigma model is
ruined by the presence of terms like ${\rm Tr } \,
[R _ { ij } R ^ { ij } ]$.
However, we have shown that these terms are not a problem
in three dimensions.

c) We found that the single loop graph was insufficient
to generate self-consistent equations of motion 
for the background fields. This was surprising, since
setting $\beta = 0$ in the usual string formalism yields
self-consistent equations of motions at the first loop level.

d) Since the formalism we have developed works for
arbitrary loop level, we can calculate higher order
corrections to the equations of motion.
We find, at two loop level, new terms which 
have the form: $R ^ 2$, $ D R  $, $F DD F $, 
$F R F $, etc.

e) We found that, by naive counting arguments on 
arbitrary loop graphs, we could absorb the leading
divergences into a rescaling of the
coupling constant $\alpha$.
By setting $\alpha \rightarrow \alpha _ R / \Lambda$, where
$\Lambda$ is a large momentum cut-off parameter, we could
absorb the leading divergences. The amplitudes
then diverge as $\Lambda ^ { 3-N}$, so the leading divergences
actually vanish if $N = 3,4,...$. For the case $N=2$, we set
the divergence to zero, thereby yielding the
equations of motions.
We found that the counting of divergences remains the same if
the coupling constant $\beta$ for the 
anti-symmetric fields is finite.
This doesn't mean that
the action is renormalizable, of course, since we still
have to analyze non-leading graphs and many other subtle
problems.

One weakness of this formalism is that 
the equations of motion emerge only after setting one
divergences to zero by hand.
Hence, we have to define the theory 
by placing in
the background fields on-shell.

In the superstring case, conformal symmetry allows
us to set $\beta = 0$. However, the entire motivation of
our approach is to analyze the D = 11 supermembrane,
where supersymmetry is sufficient to set these lower
order divergences to zero. Our ultimate goal, therefore,
is to see whether supersymmetry is strong enough to
control the divergences found in the supermembrane theory,
and whether we can obtain the
the M-theory action by expanding around higher order corrections to the 
standard D=11 supergravity action, and then use recursion relations
to probe the entire action. The supersymmetric case will be discussed
in a forthcoming paper.

\end{document}